# Temperature-dependent Raman study of CeFeAsO$_{0.9}$F$_{0.1}$ Superconductor: Crystal field excitations, phonons and their coupling

Pradeep Kumar[1], Anil Kumar[2], Surajit Saha[1], D. V. S. Muthu[1], J. Prakash[3], U. V. Waghmare[2], A. K. Ganguli[3] and A. K. Sood[1,*]

[1]Department of Physics, Indian Institute of Science, Bangalore -560012, India

[2]Theoretical Sciences Unit, Jawaharlal Nehru Centre for Advanced Scientific Research, Bangalore -560064, India

[3]Department of Chemistry, Indian Institute of Technology, New Delhi -110016, India

**ABSTRACT**

We report temperature-dependent Raman spectra of CeFeAsO$_{0.9}$F$_{0.1}$ from 4 K to 300 K in spectral range of 60 to 1800 cm$^{-1}$ and interpret them using estimates of phonon frequencies obtained from first-principles density functional calculations. We find evidence for a strong coupling between the phonons and crystal field excitations; in particular Ce$^{3+}$ crystal field excitation at 432 cm$^{-1}$ couples strongly with E$_g$ oxygen vibration at 389 cm$^{-1}$. Below the superconducting transition temperature, the phonon mode near 280 cm$^{-1}$ shows softening, signaling its coupling with the superconducting gap. The ratio of the superconducting gap to $T_c$ thus estimated to be ~ 10 suggests CeFeAsO$_{0.9}$F$_{0.1}$ as a strong coupling superconductor. In addition, two high frequency modes observed at 1342 cm$^{-1}$ and 1600 cm$^{-1}$ are attributed to electronic Raman scattering from ($x^2$-$y^2$) to $xz$ / $yz$ d-orbitals of Fe.

# 1. INTRODUCTION

The recent discovery of superconductivity in $RFeAsO_{1-x}F_x$ (R = La, Sm, Ce, Nd, Pr and Gd ) [1-3] with a transition temperature ($T_c$) varying from 26 K to 55 K has generated enormous interest because these materials are the first non-copper oxide superconductors with relatively high $T_c$. Similar to cuprates, superconductivity in rare earth Fe based oxide systems $RFeAsO_{1-x}F_x$ and $(Ba_{1-x}K_x)Fe_2As_2$ (termed as 1111 and 122 systems, respectively) is also derived from either electron [1-4] or hole [5, 6] doping of their parent compounds. The parent compound LaFeAsO exhibits long-range antiferromagnetic (AFM) order as in cuprates which is suppressed on doping [7]. The microscopic mechanism of superconductivity in iron pnictides remains an open question. The calculated electron-phonon coupling is found to be weak to produce a superconducting state within the Eliashberg theory [8], and, therefore an unconventional origin of superconductivity mediated by AFM spin fluctuation has been suggested [9]. However Boeri et al [10] have recently calculated the effect of magnetism and doping on the electron-phonon coupling and found that electron-phonon coupling increases significantly, which is strong enough to have a non-negligible effect on superconductivity. Report by Egami et. al [11] suggests that electron-phonon coupling can play crucial role relevant to the mechanism of superconductivity in iron pnictides through the spin-channel. Recently, inverse iron isotope effect on the transition temperature of $(Ba,K)Fe_2As_2$ superconductors has been observed, suggesting a competition between the conventional phonon mediated mechanism and unconventional one like antiferromagnetic fluctuation mediated mechanism [12].

In the context of high $T_c$ cuprate superconductors, Raman spectroscopy has been shown to be a powerful probe of (i) crystal field (CF) excitations and their coupling with phonons (ii) the renormalization of the phonon modes due to the opening of the superconducting gap (iii) Magnons in parent compounds and (iv) Electronic Raman background [13-14]. Room temperature Raman studies on $RFeAsO_{1-x}F_x$ (R = Ce, La, Nd and Sm) [15-19] assigned the observed phonons according to the symmetry based irreducible representations. A few temperature dependent Raman studies have been reported on $NdFeAsO_{1-x}F_x$ [16, 18], $Sr_{1-x}K_xFe_2As_2$ (x = 0 and 0.4) [20], $CaFe_2As_2$ [21], $R_{1-x}K_xFe_2As_2$ (R = Ba, Sr) [22, 23], $Ba(Fe_{1-x}Co_x)_2As_2$ [24] and $FeSe_{0.82}$ [25]. In case of $NdFeAsO_{1-x}F_x$ [16, 18] none of the observed phonon modes show any anomaly below the superconducting transition temperature. Similarly, no phonon anomaly was seen as a function of temperature in $Sr_{1-x}K_xFe_2As_2$ [20]. However, in another study of $R_{1-x}K_xFe_2As_2$ (R= Ba, Sr) [22] linewidths of the phonon modes involving Fe and As near 185 $cm^{-1}$ ($A_{1g}$) and 210 $cm^{-1}$ ($B_{1g}$) show a significant decrease below the spin-density-wave transition temperature $T_s \sim 150$ K, attributed to spin-density-wave gap opening. Also, the frequency of the 185 $cm^{-1}$ mode shows a discontinuous change at $T_s$, signaling first order structural transition accompanying the spin-density-wave transition at $T_s$. Similar results for the $B_{1g}$ mode (near 210 $cm^{-1}$) are seen for $Sr_{0.85}K_{0.15}Fe_2As_2$ and $Ba_{0.72}K_{0.28}Fe_2As_2$ ($T_s \sim 140$ K) [23]. In parent compound $CaFe_2As_2$, the $B_{1g}$ phonon frequency (210 $cm^{-1}$) shows a discontinuous decrease at $T_s \sim 173$ K and the $A_{1g}$ phonon (near 190 $cm^{-1}$) intensity is zero above $T_s$, attributed to the first order structural phase transition and a drastic change of charge distribution within the FeAs plane [21]. The $E_g$ (Fe, As) phonon (~ 135 $cm^{-1}$) in $Ba(Fe_{1-x}Co_x)_2As_2$ (x < 0.6) splits into two modes near the structural transition temperature ($T_s$ ~ 100 to 130 K) linked to strong spin-phonon coupling [24]. Similarly, $E_g$ (~ 100 cm$^{-1}$) mode in case of FeSe$_{0.82}$ associated with the Se vibrations in *ab* plane shows anomalous hardening below $T_s$ (~ 100 K) attributed to the strong spin-phonon coupling [25]. Also electronic Raman scattering has been reported in FeSe$_{0.82}$ associated with the Fe *d*-orbitals [25].

However, there are no detailed temperature dependent Raman studies on parent or doped CeFeAsO. In this paper we report such a study of CeFeAsO$_{0.9}$F$_{0.1}$ with a superconducting transition temperature $T_c$ ~ 38 K [26]. There are two motivating factors behind this work: first, Raman study of phonons in these materials can provide information on the superconducting state through coupling of phonons to Raman active electronic excitations. Second, electronic excitations between the crystal-field split *f*-orbitals of Ce$^{3+}$ and *d*-orbitals of Fe can be probed. This is important because Fe orbitals contribute significantly to the electronic states at the Fermi level and hence they are expected to play a crucial role in the mechanism of superconductivity. In particular, it is not clear experimentally if $d_{xz}$ and $d_{yz}$ orbitals are split or not [27-30] in the general class of iron based superconductors. In ref [29], related to LaFeAsO, the splitting of *3d* orbitals of Fe results from combined effects of tetrahedral crystal field, spin-orbit coupling, strong hybridization between Fe *d* and As *4p* orbitals and lattice compression along *z*-axis. Also, it has been suggested [30] that, since $d_{xz}$ and $d_{yz}$ orbitals are roughly half filled, spontaneous symmetry breaking can lead to the lifting of the degeneracy of $d_{xz}$ and $d_{yz}$ orbitals.

Here, from the temperature dependence of Raman scattering from phonons and electrons (crystal-field split *d*-orbitals of Fe) in tetragonal- CeFeAsO$_{0.9}$F$_{0.1}$, we present four significant results: (i) a phonon mode at 389 cm$^{-1}$ associated with the oxygen vibrations shows strong coupling with $Ce^{3+}$ $f$-level crystal field excitation observed at 432 cm$^{-1}$ (ii) Raman scattering from $Ce^{3+}$ $f$-levels and higher order Raman scattering from combination of the CF and phonon excitations is seen. (iii) A phonon mode near 280 cm$^{-1}$ shows softening below $T_c$, which can arise from coupling between the superconducting gap $2\Delta$ and the phonon, thereby giving $2\Delta/K_B T_c \sim 10$. (iv) High frequency Raman modes are observed at 1342 cm$^{-1}$ and 1600 cm$^{-1}$, attributed to electronic Raman scattering between $x^2$-$y^2$ and $xz/yz$ $d$-orbitals. This implies that $d_{xz}$ and $d_{yz}$ orbitals of Fe are non-degenerate with a splitting of ~ 32 meV.

## 2. METHODS

### 2.1 Experimental Details

Polycrystalline samples of $CeFeAsO_{0.9}F_{0.1}$ with a superconducting transition temperature of 38 K and without any impurity phase were prepared and characterized as described in ref. 26. Unpolarised micro-Raman measurements were performed in backscattering geometry, using 514.5 nm line of an Ar-ion Laser (Coherent Innova 300) and Raman spectrometer (DILOR XY) coupled to a liquid nitrogen cooled CCD detector. Temperature variation was done from 4 K to 300 K, with a temperature accuracy of ± 0.1K, using continuous flow He cryostat (Oxford Instrument, ITC502).

### 2.2 Computational Details

Our first-principles calculations are based on density functional theory as implemented in the PWSCF [31] package. We use optimized norm-conserving pseudopotential for Ce [32, 33] constructed with $Ce^{+3}$ as a reference state and ultrasoft pseudopotentials [34] for other elements (O, Fe, As) to describe the interaction between ionic cores and valence electrons, and exchange correlation energy functional with a local density approximation. We use plane wave basis with a kinetic energy cutoff of 60 Ry in representation of wavefunctions and a cutoff of 360 Ry for representing the charge density. Convergence with respect to basis set and grid used in the Brillouin zone sampling has been carefully checked. We sampled integration over the Brillouin zone (of single unit cell) with 12×12×6 Monkhorst Pack Mesh [35]. For Fluorine-substituted compound, we used a $\sqrt{2}x\sqrt{2}x1$ supercell. Structural optimizations of CeFeAsO and CeFeAsO$_{0.75}$F$_{0.25}$ are carried through minimization of the total energy using Hellman-Feynman forces and the Broyden-Flecher-Goldfarb-Shanno based method. Zone center (q = (0,0,0)) phonon spectra are determined using a frozen phonon method (with atomic displacements of 0.04 Å) for the relaxed structure obtained at experimental lattice constants. In addition, we used DFT-linear response [31] to obtain phonon spectra at wave-vectors on a 2x2x2 mesh and Fourier interpolation of the interatomic force constants to determine phonons at wave-vectors on a finer mesh (8x8x8), and obtain one- and two-phonon density of states.

## 3. RESULTS AND DISCUSSIONS

### 3.1. Raman Scattering from Phonons

CeFeAsO has a layered structure belonging to the tetragonal *P4/nmm* space group containing two CeFeAsO units per unit cell. There are eight Raman active phonon modes belonging to the irreducible representation $2A_{1g} + 2B_{1g} + 4E_g$, with the dominant displacement of atoms classified as La: $A_{1g} + E_g$ ; O/F: $B_{1g} + E_g$ ; Fe: $B_{1g} + E_g$ and As: $A_{1g} + E_g$ [17]. Figure 1 shows Raman spectrum at 4 K, revealing 14 modes labeled as S1 to S14 in the spectral range of 60-1800 cm$^{-1}$. Raman spectra at a few temperatures are shown in frequency range 200 to 600 cm$^{-1}$ in Fig.2 and from 600 to 1750 cm$^{-1}$ in Fig. 3. Spectra are fitted to a sum of Lorentzian functions. The individual modes are shown by thin lines and resultant fit by thick lines. It can be seen from Fig. 2 that the mode near 400 cm$^{-1}$ is highly asymmetric and can be fitted well to a sum of two Lorentzian functions, representing modes S5 and S6. The frequencies measured at 4 K are tabulated in Table I.

A noticeable feature of our data is observation of modes S8 to S14 above 600 cm$^{-1}$. The available experiments [15-25] and theoretical calculation [36] of phonons in the iron based superconductors show that first order Raman phonons are observed only below 550 cm$^{-1}$. Before we discuss assignment of modes S1 to S5 and mode S7 as phonon modes, we review the assignments of the Raman modes observed so far in (1111) systems [15-19] to help us in associating the observed modes given in Table I. Out of the eight Raman active modes, three distinct modes observed at room temperature in CeFeAsO$_{0.84}$F$_{0.16}$ are 220 cm$^{-1}$ (B$_{1g}$, Fe), 280 cm$^{-1}$ (E$_g$, Fe) and 450 cm$^{-1}$ (E$_g$, O) [15]. In addition, a weak mode could be seen at 396 cm$^{-1}$ as a shoulder of the 450 cm$^{-1}$ mode [15]. In NdFeAsO$_{0.9}$F$_{0.1}$, five Raman active modes have been observed and assigned as 252.5 cm$^{-1}$ (mixed mode of Fe and As in *c*-direction), 262.5 cm$^{-1}$ (mixed mode of Fe and As in *a* or *b* direction), 337.8 cm$^{-1}$ (oxygen mode), 368 cm$^{-1}$ (Nd-O or Nd-F vibration) and 487 cm$^{-1}$ (Nd-O stretching mode) [16]. In LaFeAsO system, six observed modes have been assigned as 96 cm$^{-1}$ (E$_g$, La), 137 cm$^{-1}$ (E$_g$, As and Fe), 161 cm$^{-1}$ (A$_{1g}$, La), 214 cm$^{-1}$ (B$_{1g}$, Fe), 278 cm$^{-1}$ (E$_g$, Fe) and 423 cm$^{-1}$ (E$_g$, O) [15]. In SmFeAsO four Raman modes have been identified as 170 cm$^{-1}$ (A$_{1g}$, Sm), 201 cm$^{-1}$ (A$_{1g}$, As), 208 cm$^{-1}$ (B$_{1g}$, Fe) and 345 cm$^{-1}$ (B$_{1g}$, O) [17]. The calculated value of (E$_g$,O) phonon frequency for SmFeAsO is 503 cm$^{-1}$ [19].

Keeping these data in view and our density functional calculations (discussed below), we assign the modes S1 to S5 to be the first order Raman active modes as S1 (102 cm$^{-1}$, $E_g$, Ce ), S2 (162 cm$^{-1}$, $A_{1g}$, Ce), S3 (223 cm$^{-1}$, $B_{1g}$, Fe), S4 (281 cm$^{-1}$, $E_g$, Fe ) and S5 (389 cm$^{-1}$, $E_g$, O) (see Table I). We have calculated phonons at Γ- point (q = 0, 0, 0) of CeFeAsO$_{1-x}$F$_x$ for x = 0.00 and x = 0.25 using first-principles density functional theory described in section 2.2. Calculated phonon frequencies for both experimental and theoretical lattice constants are listed in Table I.

The lattice constant used are as follows, Experimental: *a* = 3.996 Å, *c* = 8.648 Å and Theoretical: *a* = 3.808 Å, *c* = 8.242 Å. Our first-principles theoretical estimates of the optimized lattice constants are 4-5% smaller than the experimental values (as found typically for pnictides, and consistent with the earlier reports [37] ). To identify phonon modes observed in the Raman spectra, we calculate phonons (using both experimental and theoretical lattice constants) at gamma point (q = (0,0,0)) of CeFeAsO$_{1-x}$F$_x$ for x = 0.00 and x = 0.25 using frozen phonon method. Calculated phonon frequencies at the experimental lattice constants are in fairly good agreement with the experimental values (see Table 1) and corresponding eigen modes are shown in Fig 4. Generally, we find that modes are localized in Ce-O and Fe-As layers. We find that the Raman-active $E_g$ modes with Ce and O character change notably with F-concentration *x*, while $E_g$ (Fe) mode is quite insensitive to *x*. A similar trend is found for the $B_{1g}$ modes. This is expected as Ce is coordinated with oxygen while Fe is coordinated with As. Secondly, all the phonon frequencies increase by 10-25 % with reduction in the lattice constant by about 5 % (from the experimental value to the LDA value). This gives us some idea about the errors involved in the theoretical estimates of phonon frequencies. In order to explore the possibility of second order Raman scattering, we also determine (using experimental lattice constants) one phonon density of states ($D(\omega)=\sum_i \delta(\omega-\omega_i)$) and two-phonon density of states (DOS) ($D(\omega)=\sum_{i,j}\delta(\omega-\omega_i+\omega_j)$) by calculating phonons at 2×2×2 grid of q-points using linear response method (see Fig 7). We find peaks at ~560 and 700 cm$^{-1}$ that correlate with S7 and S8 modes observed in Raman spectra. Based on our estimates, the modes above 800 cm$^{-1}$, are likely to arise from the electronic Raman scattering, such as ones associated with crystal field splitting of Fe $d$-levels.

**3.2 Raman Scattering from Crystal Field Split Excitations of $Ce^{3+}$**

We now discuss why we assign S6 mode ( ~ 435 cm$^{-1}$) as electronic Raman scattering involving the crystal field split excitations of $Ce^{3+}$ J-levels (J = 5/2). Inelastic neutron scattering from $CeFeAsO_{1-x}F_x$ (x = 0, 0.16) shows that the $Ce^{3+}$ crystal field levels are three magnetic doublets at 0 ($E_0$), 18.7 meV ($E_1$) and 58.4 meV ($E_2$) [38], thereby allowing three possible transitions: $E_0$ to $E_1$ (~ 150 cm$^{-1}$), $E_0$ to $E_2$ (~ 470 cm$^{-1}$) and $E_1$ to $E_2$ (~ 320 cm$^{-1}$ ) [38]. At low temperatures only the first two transitions can be seen. Raman cross section for the crystal field excitation (CFE) is often weak and these have been observed mixed with nearby Raman-active phonons via electron-phonon interaction [39]. The frequency of the mode S6 is very close to the transition from ground state ($E_0$) to the second excited state ($E_2$). Moreover, the integrated intensity of the S6 mode decreases with increasing temperature (shown in Fig.5a), which is typical for Raman scattering associated with crystal-field transitions (because the ground state population decreases with increasing temperature). We, therefore, assign S6 to electronic Raman scattering from the crystal filed split levels of $Ce^{3+}$. The transition from $E_0$ to $E_1$ is not clearly seen probably because of its overlap with Raman phonon S2.

A coupling between CFE and phonon mode is expected only if they have comparable energies and are of the same symmetry [39]. Since $Ce^{3+}$ ion is located between planes of oxygen and arsenic, phonon modes involving oxygen and arsenic will be expected to couple strongly to the CFE via their modulation of the crystal field. The electric field due to surrounding ions split the $Ce^{3+}$ ion $^2F_{5/2}$ multiplets into one $M^{6-}$ and two $M^{7-}$ (Koster notation [40]) Kramers doublets, denoted by $E_0$ ($M^{6-}$), $E_1$ ($M^{7-}$) and $E_2$ ($M^{7-}$). The symmetry of the CF excitation within this multiplet is given by the direct product of the irreducible representation of the doublets: $E_0$ to $E_1$ or $E_2$: $M^{6-} \otimes M^{7-} = B_{1g} + B_{2g} + E_g$ and $E_1$ to $E_2$ : $M^{7-} \otimes M^{7-} = A_{1g} + A_{2g} + E_g$. Therefore, the CF excitation from the ground state ($E_0$) to second excited state ($E_2$) is group theoretically allowed to couple with the mode S5 ($E_g$).

Figure 5(b) shows the temperature dependence of full width at half maxima (FWHM) of modes S5 and S6. It can be seen that FWHM of S5 mode is independent of temperature whereas that of S6 is anomalous, in the sense that the linewidth increases on decreasing temperature. The frequency of mode S6 remains nearly independent of temperature (Fig 6). Similar anomalies have been reported in cuprates for the crystal field mode [41]. These observations suggest that $Ce^{3+}$ CF mode (S6) is coupled strongly with the phonon mode at 389 cm$^{-1}$ (S5) associated with the oxygen vibrations.

### 3.3. Temperature Dependence of the Phonon Frequencies

Mode S1, S2 and S3 are weak and their temperature dependence is difficult to quantify; we, therefore, focus on the temperature dependence of peak positions of S4 to S7 and S9, as shown in Fig 6. The solid line for mode S5 in Fig.6 is fitted to an expression based on cubic anharmonicity where the phonon mode decays into two equal energy phonons [42]:

ω (T) = ω (0) + C [1+2n (ω(0)/2)], where ω (T) and ω (0) are the phonon frequencies at temperature T and T = 0, C is Self-energy parameter for a given phonon mode and n(ω) = 1/[exp($\hbar\omega/\kappa_B T$) -1] is the Bose-Einstein mean occupation factor. Fitted parameters for the mode S5 are ω (0) = 391.14 cm$^{-1}$, C = -5.12 ± 0.44 cm$^{-1}$.

Phonons which couple strongly to electronic states near Fermi-surface can be easily influenced by changes in the neighborhood of the Fermi-surface. In superconductors, the opening of the superconducting gap redistributes the electronic states in the neighborhood of the Fermi-surface and hence will change phonon self-energy seen in cuprate superconductors [43]. Qualitatively, phonons below the gap show a softening whereas phonons with frequency above the gap value can harden. The latter can turn into a small softening if impurity scattering is taken into account [44]. The anomalous softening of the S4 mode below $T_c$ (shown in Fig. 6) indicates a coupling of this phonon to the electronic system. As superconducting gap pushes the phonon towards lower energy below $T_c$, we suggest that superconducting gap in this system should lie close to this phonon towards high energy side i.e. *2Δ* ≥ 281 cm$^{-1}$, similar to a softening of the Raman active phonon mode below $T_c$ in cuprate superconductors [43]. This gives an estimate of the ratio of gap (*2Δ*) to the transition temperature *2Δ/K$_B$T$_c$* ~ *10*. This ratio suggests that this system belongs to the class of strong-coupling superconductors. We note that the present value of *2Δ/K$_B$T$_c$* is close to the value found for (Nd, Sm, La)FeAsO$_{1-x}$F$_x$ (*2Δ/K$_B$T$_c$* ~ *8*) [45-47] and also ( SrBa)$_{1-x}$(K, Na)$_x$Fe$_2$As$_2$ (*2Δ/K$_B$T$_c$* ~ *8*) [48]. There are reports of multiple gap in (1111 and 122) systems with values of *2Δ/K$_B$T$_c$* ranging from 3 to 9 [47, 49-50].

**3.4. Origin of the High Frequency Modes S7, S8, S9, S10, S11 and S12**

In addition to the expected first-order Raman bands, we observe modes S7 to S14. A weak mode was observed at the frequency close to that of mode S7 in the room temperature Raman study of $CeFeAsO_{1-x}F_x$ [15], attributed to multiphonon Raman scattering. We make similar assignment for the mode S7. A broad mode around 720 cm$^{-1}$ (S8) is similar to a broad band in Raman spectra of superconductor $MgBr_2$ [14], which has been attributed to a maximum in density of states (DOS) arising from the disorder induced contributions from phonons away from the zone centre.

The coupled CFE with the longitudinal optical phonons have been observed in case of $UO_2$ [51]. Following this, the mode S9 (855 cm$^{-1}$) is assigned as a second order Raman scattering from a combination of modes S5 and S6. The weak modes S10, S11 and S12 can be multiphonon Raman scattering.

### 3.5. Electronic Raman Scattering from Fe 3d-Orbitals

Two high energy excitations at 1342 cm$^{-1}$ (S13) and 1600 cm$^{-1}$ (S14) are observed in our experiments. In earlier Raman study of $CeFeAsO_{1-x}F_x$ [15] at room temperature, weak Raman modes were observed at 846 and 1300 cm$^{-1}$ and it was suggested [15] that the high frequency modes may be related to the electronic Raman scattering involving the *d*-orbitals of Fe. Our recent Raman studies on $FeSe_{0.82}$ superconductors have also shown similar modes and were attributed to electronic Raman scattering between crystal field split *d*-orbitals of Fe as $x^2-y^2$ to *xz* and *yz* [25]. We make this assignment in the present case as well, pointing out a splitting of *d*-orbitals *xz* and *yz* as ~ 32meV.

### 4. CONCLUSIONS

In conclusion, we have presented Raman measurements of $CeFeAsO_{0.9}F_{0.1}$ as a function of temperature. We suggest that the softening of the Raman active phonon mode (S4) below $T_c$ is due to an opening of a superconducting gap, yielding $2\Delta/K_B T_c \sim 10$. A Raman mode at ~ 432 cm$^{-1}$ is attributed to electronic Raman scattering involving $Ce^{3+}$ crystal field levels, coupled strongly to $E_g$ oxygen phonons. A strong mode is observed at 855 cm$^{-1}$ which has been attributed to a combination mode of the above crystal field excitation and the $E_g$ oxygen vibration. High frequency modes observed at 1342 cm$^{-1}$ and 1600 cm$^{-1}$ are attributed to electronic Raman scattering involving $x^2$-$y^2$ to ($xz$, $yz$) d-orbitals of Fe.

## Acknowledgments

AKS and AKG acknowledge the DST, India, for financial support. PK, AK and JP acknowledge CSIR, India, for research fellowship. UVW acknowledge DAE Outstanding Researcher Fellowship for partial financial support.

Table-I: List of the experimental observed frequencies at 4K and calculated frequencies in $CeFeAsO_{1-x}F_x$.

| Mode Assignment | Experimental ω (cm$^{-1}$) | Calculated ω (cm$^{-1}$) | | | |
| --- | --- | --- | --- | --- | --- |
| | | Experimental Lattice constant | | Theoretical Lattice constant | |
| | | x = 0.0 | x = 0.25 | x = 0.0 | x = 0.25 |
| S1  $E_g$ (Ce) | 102 | 101 | 90 | 127 | 112 |
| S2  $A_{1g}$ (Ce) | 162 | 171 | 162 | 200 | 189 |
| S3  $B_{1g}$ (Fe) | 223 | 201 | 202 | 233 | 241 |
| S4  $E_g$ (Fe) | 281 | 287 | 285 | 323 | 322 |
| S5  $E_g$ (O) | 389 | 394 | 417 | 467 | 486 |
| S6  CF of $Ce^{3+}$ levels | 432 | | | | |
| S7  Two-Phonon | 524 | | | | |
| S8  DOS | 720 | | | | |
| S9  (S5 + S6) | 855 | | | | |
| S10 Multiphonon | 932 | | | | |
| S11 Multiphonon | 975 | | | | |
| S12 Multiphonon | 1026 | | | | |
| S13 CF of Fe-d levels | 1342 | | | | |
| S14 CF of Fe-d levels | 1600 | | | | |

**FIGURE CAPTION**

FIG.1. Unpolarised Raman spectra of $CeFeAsO_{0.9}F_{0.1}$ at 4 K.

FIG.2. (Color online) Raman spectra from 200 to 600 $cm^{-1}$. Solid (thin) lines are fit of individual modes and solid (thick) line shows the total fit to the experimental data (open circle).

FIG.3. (Color online) Temperature evolution of the high frequency (600 to 1750 $cm^{-1}$) modes. Solid (thin) lines are fit of individual modes and solid (thick) line shows the total fit to the experimental data (open circle).

FIG.4. (Color online) Eigen modes corresponding to different Raman modes in Table (1).

FIG.5. Temperature dependence of the Linewidth of modes S5, S6 (Top panel) and Normalized integrated intensity of S6 mode (Bottom panel).

FIG.6. (Color online) Temperature dependence of the modes S4, S5, S6, S7, and S9. Solid lines for S4, S6, S7, and S9 are drawn as guide to the eye. The solid line for S5 is the fitted curve as described in text.

FIG.7. Calculated (a) Phonon density of states and (b) Two-phonon density of states of CeFeAsO at experimental lattice constant.

Figure1:

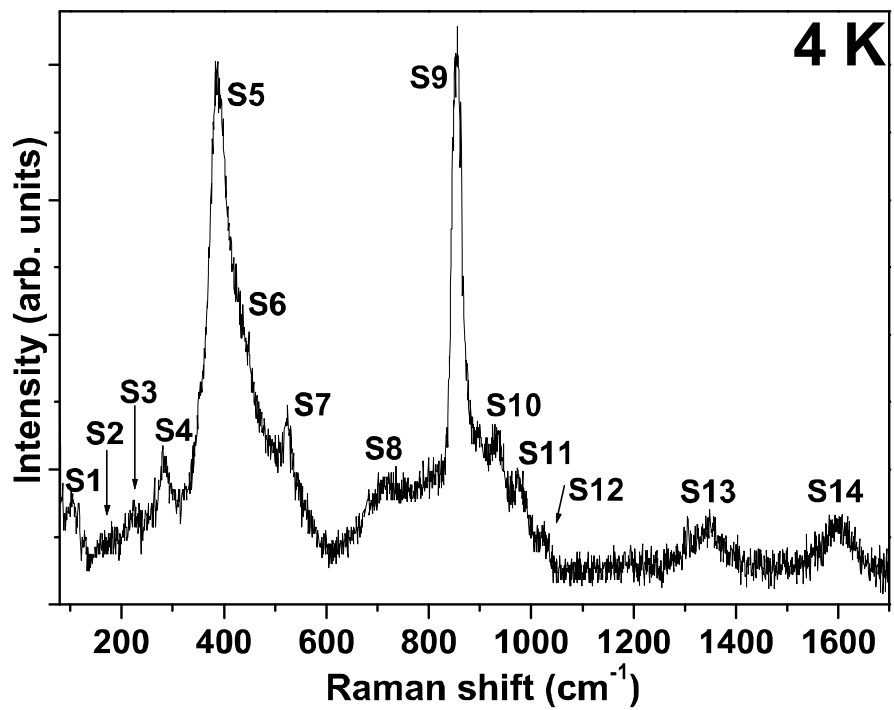

Figure2 :

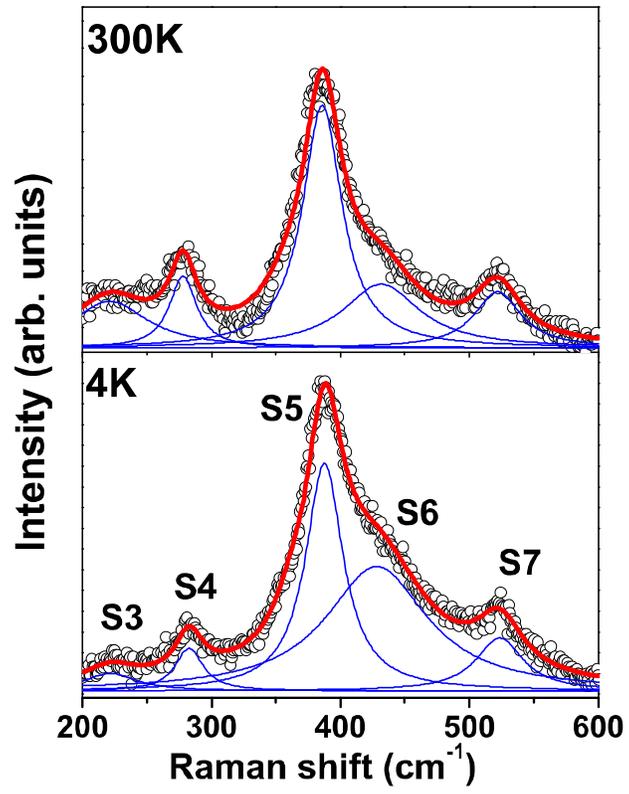

Figure3 :

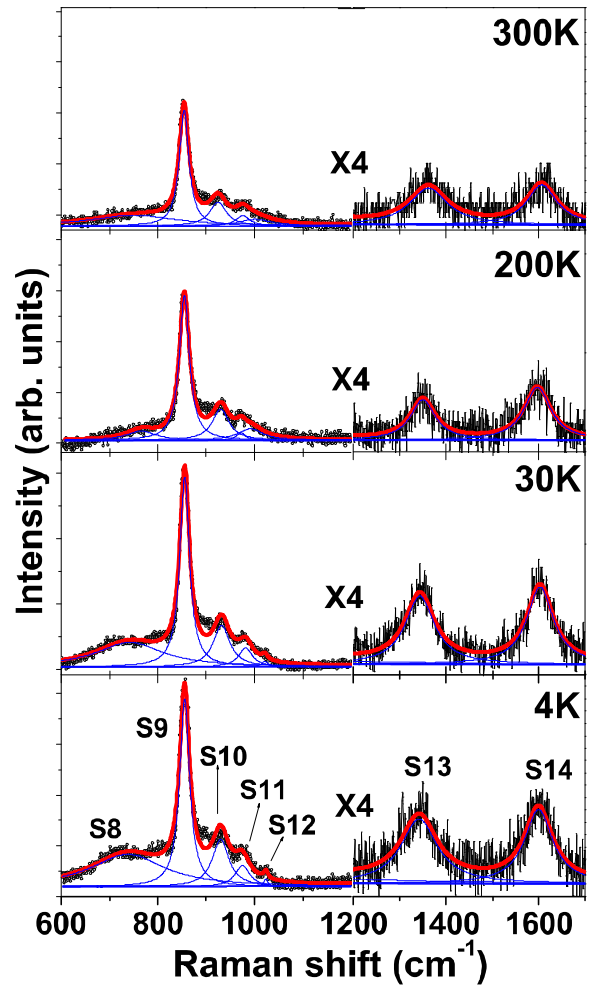

Figure4 :

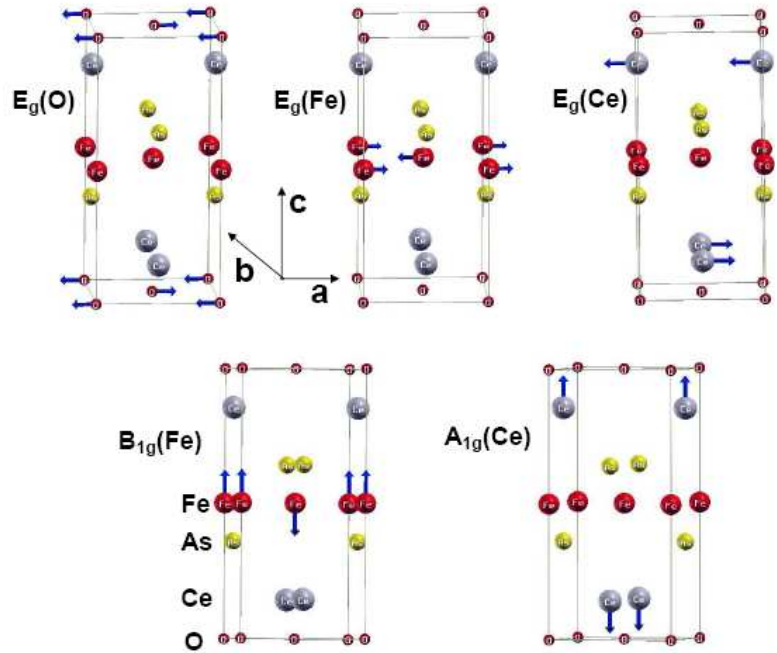

Figure5:

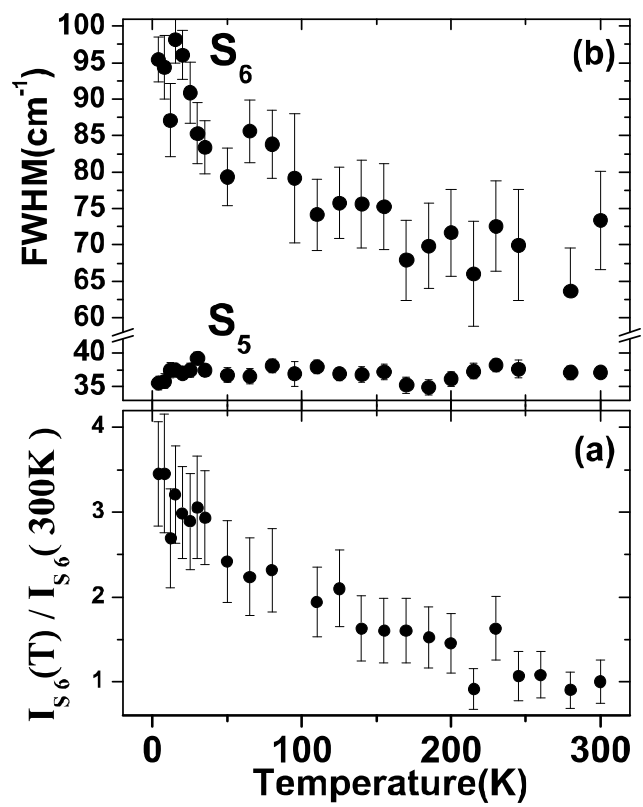

Figure6:

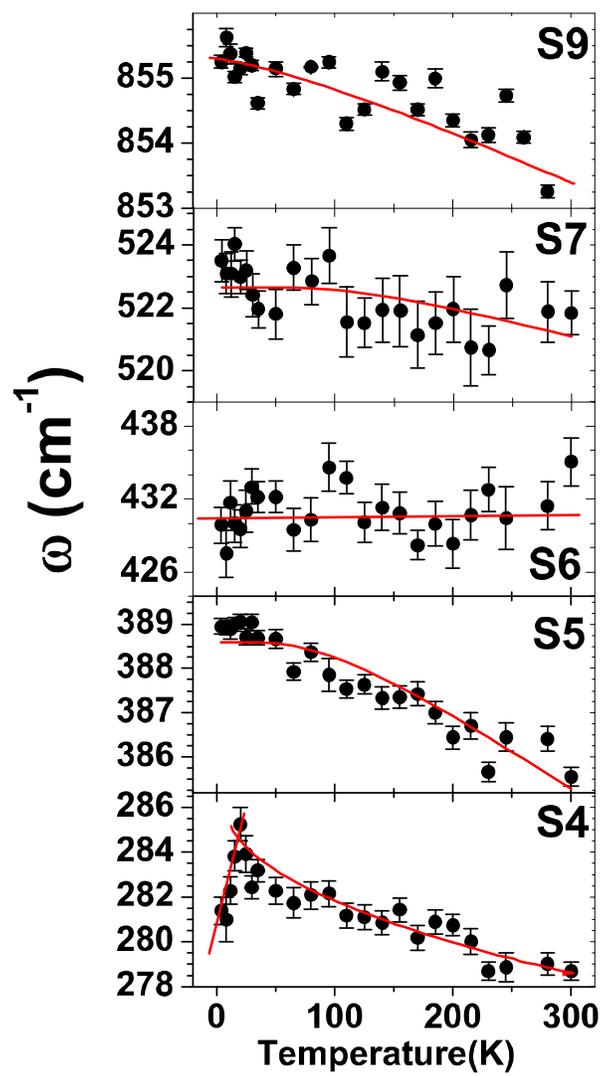

Figure 7:

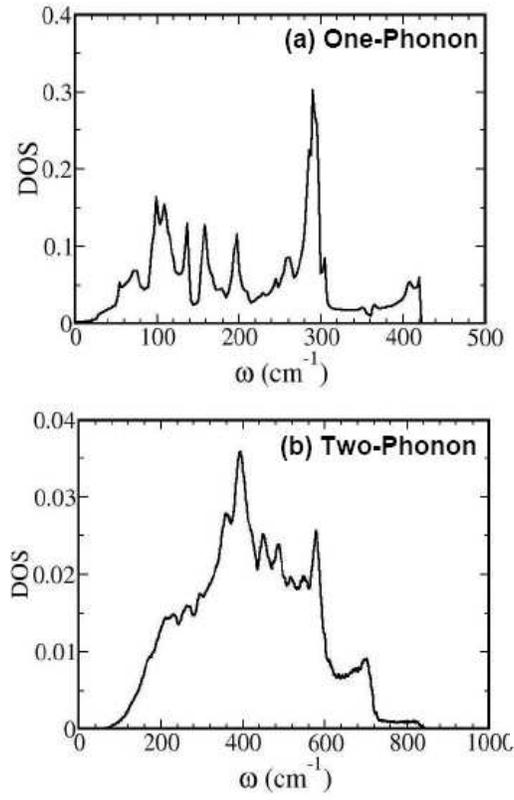